\documentclass[journal]{IEEEtran}
\usepackage[ruled,vlined]{algorithm2e}
\usepackage{graphicx} 
\usepackage{textcomp}
\usepackage[most]{tcolorbox}
\usepackage{multirow}
\usepackage{booktabs}
\usepackage{subfig}
\usepackage{bbding,cleveref}
\usepackage{wrapfig}
\usepackage{paralist}
\usepackage{xspace}
\usepackage{pifont,wasysym}

\NewDocumentCommand{\rotb}{O{45} O{1em} m}{\makebox[#2][l]{\rotatebox{#1}{#3}}}%

\newcommand{\putfw}{\texttt{putFirmware}}
\newcommand{\vetofw}{\texttt{vetoFirmware}}
\newcommand{\checkfw}{\texttt{checkFirmware}}
\newcommand{\putlog}{\texttt{putMeasurement}}

\newcommand{\proposeput}{\texttt{proposePubKey}}
\newcommand{\voteput}{\texttt{votePubKeyProposal}}

\newcommand{\updateadminkey}{\texttt{updateAdminKey}}

\newcommand{\keygen}{\mathsf{Key}$-$\mathsf{Gen}}
\newcommand{\tpmkeygen}{\mathsf{TPM.Key}$-$\mathsf{Gen}}
\newcommand{\oemkeygen}{\mathsf{OEM.Key}$-$\mathsf{Gen}}

\newcommand{\hlnkeygen}{\mathsf{HLN.Key}$-$\mathsf{Gen}}
\newcommand{\threskeygen}{\mathsf{Thres.Key}$-$\mathsf{Gen}}
\newcommand{\sign}{\mathsf{Sig}}
\newcommand{\tpmsign}{\mathsf{TPM.Sig}}
\newcommand{\oemsign}{\mathsf{OEM.Sig}}

\newcommand{\hlnsign}{\mathsf{HLN.Sig}}
\newcommand{\thressign}{\mathsf{Thres.Sig}}
\newcommand{\verify}{\mathsf{Ver}}
\newcommand{\tpmverf}{\mathsf{TPM.Ver}}
\newcommand{\oemverf}{\mathsf{OEM.Ver}}

\newcommand{\hlnverf}{\mathsf{HLN.Ver}}
\newcommand{\thresverf}{\mathsf{Thres.Ver}}
\newcommand{\pk}{\textit{pk}}
\newcommand{\tpmpk}{\textit{pk}^{\textsc{TPM}}}
\newcommand{\oempk}{\textit{pk}^{\textsc{OEM}}}

\newcommand{\hlnpk}{\textit{pk}^{\textsc{HLN}}}

\newcommand{\sk}{\textit{sk}}
\newcommand{\tpmsk}{\textit{sk}^{\textsc{TPM}}}
\newcommand{\oemsk}{\textit{sk}^{\textsc{OEM}}}

\newcommand{\hlnsk}{\textit{sk}^{\textsc{HLN}}}

\newcommand{\msg}{\textit{m}}

\newcommand{\github}{{https://github.com/FLBI-Team/flbi}}

\author{\IEEEauthorblockN{Dani\"el Reijsbergen, Aung Maw, Sarad Venugopalan, Dianshi Yang, Tien Tuan Anh Dinh, and Jianying Zhou} \\[0.1cm]
\IEEEauthorblockA{Singapore University of Technology and Design}}

\title{Protecting the Integrity of IoT Sensor Data and Firmware With A Feather-Light Blockchain Infrastructure}

\begin{document} 

\IEEEoverridecommandlockouts
\IEEEpubid{\makebox[\columnwidth]{978-1-6654-9538-7/22/\$31.00~\copyright2022 IEEE \hfill} \hspace{\columnsep}\makebox[\columnwidth]{ }}

\maketitle

\begin{abstract}
Smart cities deploy large numbers of sensors and collect a tremendous amount of data from them. For example, Advanced Metering Infrastructures (AMIs), which consist of physical meters that collect usage data about public utilities such as power and water, are an important building block in a smart city. In a typical sensor network, the measurement devices are connected through a computer network, which exposes them to cyber attacks. Furthermore, the data is centrally managed at the operator's servers, making it vulnerable to insider threats.

Our goal is to protect the integrity of data collected by large-scale sensor networks and the firmware in measurement devices from cyber attacks and insider threats.  To this end, we first develop a comprehensive threat model for attacks against data and firmware integrity, which can target any of the stakeholders in the operation of the sensor network. Next, we use our threat model to analyze existing defense mechanisms, including signature checks, remote firmware attestation, anomaly detection, and blockchain-based secure logs. However, the large size of the Trusted Computing Base and a lack of scalability limit the applicability of these existing mechanisms. We propose the Feather-Light Blockchain Infrastructure (FLBI) framework to address these limitations. Our framework leverages a two-layer architecture and cryptographic threshold signature chains to support large networks of low-capacity devices such as meters and data aggregators. We have fully implemented the FLBI's end-to-end functionality on the Hyperledger Fabric and private Ethereum blockchain platforms. Our experiments show that the FLBI is able to support millions of end devices.

\end{abstract}


\newcommand{\nmeters}{N}
\newcommand{\meastime}{t}
\newcommand{\meassize}{b}
\newcommand{\ndcus}{M}
\newcommand{\nshards}{S}

\newcommand{\daniel}[1]{\textcolor{red}{Daniel: #1}}

\section{Introduction}
\label{sec:introduction}

As part of the global Smart Cities drive, IoT sensor networks have emerged that consist of hundreds of thousands of measurement devices. 
This growth has been driven by the development of efficient IoT networking protocols \cite{alfugaha2015internet} and an abundance of low-power measurement devices.
\textit{Advanced Metering Infrastructures} (AMIs), which consist of smart meters that do not require manual readouts \cite{skopik2015smart}, are a prominent example. Meters in an AMI
take readings at high frequency and send them to a utility company over a computer network. 
Other examples of large-scale sensor networks include security cameras, of which some 200,000 are planned to be installed in Singapore by 2030 \cite{reuters2021}, and patient monitoring in a city-wide healthcare system \cite{islam2015internet}.

Meanwhile, sensor networks are also becoming valuable targets for cyber attacks. For example, AMIs are tightly coupled
with critical infrastructures such as power, gas, and water~\cite{liu2012cyber}, which attract 
powerful adversaries~\cite{ukraine2016analysis}. 
In particular, by tampering with the \textit{integrity} of {AMI data}, an attacker can deceive a utility company into making incorrect grid balancing decisions or sending incorrect bills. This can have a large-scale, visible
impact on people's lives, e.g., attackers can cause wide-area
blackouts and cascading power line failures~\cite{soltan2018blackiot,shekari2021mamiot}. 
Similarly, attacks against security cameras can impair crime-fighting efforts, whereas attacks against healthcare data can endanger the lives of patients.
An additional AMI security threat involves the integrity of the \textit{firmware} on the measurement devices: by overwriting the firmware, an attacker can seize control of the device, as witnessed by the recent Mirai, VPNFilter, and Prowli botnets \cite{zheng2019firm}.
Attacks against data or firmware integrity can remain undetected until the adverse effects -- e.g., incorrect load balancing decisions or a botnet attack -- have become evident. Detecting these attacks in a timely manner is therefore an important security challenge.

Many solutions for securing data and firmware integrity have been proposed for IoT networks \cite{stellios2018survey}, mobile networks, and distributed systems. 
As a second contribution, we design a blockchain framework, called the Feather-Light Blockchain Infrastructure (FLBI), that addresses the above limitations. 
The FLBI framework allows a consortium of operators and other interested parties to maintain a secure log of measurement and firmware updates. 
Our framework addresses the challenge of minimizing the TCB  through the use of a \textit{signature chain} that is maintained by the evolving consortium. The signature chain is verified inside the device \textit{bootloader}. We keep the added complexity minimal through the use of \textit{threshold signatures} \cite{gennaro2018fast}, which shift the computational burden to the operators' servers. To the best of our knowledge, this is the first application of chains of threshold signatures in this context. The FLBI addresses the blockchain scalability problem by using a \textit{two-layer blockchain architecture} \cite{gudgeon2020sok}. This architecture consists of a top-layer blockchain maintained by the consortium members, and a large number of bottom-layer blockchains running on low-capacity devices that are in close proximity to the devices (e.g., data concentrators). Although two-layer mechanisms have been proposed before in the literature, they often lack a concrete, publicly accessible implementation. 
Furthermore, the FLBI includes mechanisms that defend against a variety of blockchain-specific attacks. In particular, it distributes nodes evenly across bottom-layer blockchains, introduces cooldown periods, and supports round-robin leaders.

As a third contribution, we have developed end-to-end implementations of the FLBI on both Hyperledger Fabric and private Ethereum. We have evaluated their performance on a testbed consisting of Raspberry Pis that represent bottom-layer blockchain nodes, and an IoT development board with a modified bootloader that represents an end device. By contrast, many existing blockchain-based approaches for data integrity lack realistic end-to-end experiments \cite{LiangWLZD19,lee2017blockchain,YohanL18}. Our experiments show that the FLBI is able to support millions of devices. The code that we used for our experiments is publicly available at \github. 

In summary, we make the following  contributions:

\begin{enumerate}
\item A comprehensive threat model for data and firmware integrity that includes attacks against all the main entities in an IoT sensor network -- we survey existing defense mechanisms and evaluate them within this model.
\item A blockchain framework for IoT data and firmware integrity that minimizes the TCB through the use of threshold signature chains, and which achieves scalability through a two-layer architecture.
\item An end-to-end implementation of our framework with real-world smart meters and an IoT development board. Our experiments show that
our framework can support hundreds of thousands of end devices. 
\end{enumerate}

Throughout this work, we will focus on a single case study, namely \textit{smart grids and AMIs}, for illustrative purposes.

The structure of this paper is as follows. We first give an overview of the relevant background on smart grids, blockchains, and digital signatures in \Cref{sec:background}. We present our system model and requirements in \Cref{sec:system_model}, and the corresponding threat model in \Cref{sec:threat_model}. We discuss the limitations of existing methods in \Cref{sec:existing_countermeasures}, and present the FLBI framework that addresses  them in \Cref{sec:architecture}. 
We present our empirical results in \Cref{sec:evaluation}, discuss related work in \Cref{sec:related_work}, and conclude the paper in \Cref{sec:conclusions}.

\section{Background}
\label{sec:background}

\textit{Smart Meters. }
As mentioned in the introduction, we will focus in this work on AMIs as an illustrative example of large-scale sensor networks.
The main building block in an AMI is the \textit{smart meter}, whose purpose is to convert a physical \textit{input signal} -- e.g., power or water flow -- into a digital \textit{measurement}. The main components of the smart meter include the sensor, a microcontroller unit (MCU) that runs the device firmware, memory (flash memory and RAM), and a communication interface \cite{KumarLBPDM19,sharma2015performance}. 
Other components typically include the power supply, an analog-to-digital converter (ADC), a real-time clock, and an LCD display 

\begin{figure}[tb]
    \centering
    \includegraphics[width=0.4\textwidth]{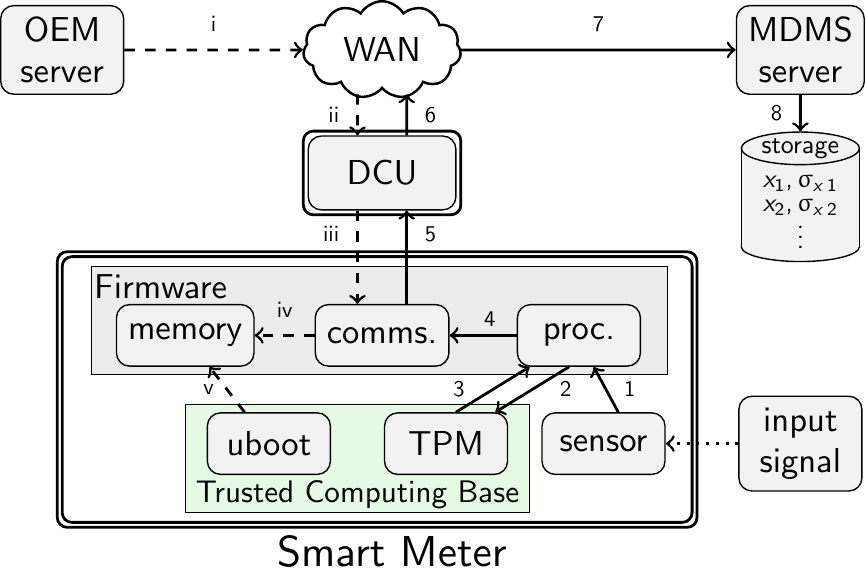}
    \caption{AMI system model.}
    \label{fig:system_model}
\end{figure}

The MCU is responsible for processing information from the sensors, whereas the communication interface allows the meter to interact with its operator (i.e., the utility company).
The measurements are stored on the MCU's built-in flash memory until they are pulled by the operator. To enhance security, a meter may also include secure components such as a Trusted Platform Module (TPM) that can be used to sign messages or for remote firmware attestation \cite{sailer2004design,kil2009remote,lemay2012cumulative}. Another secure component is the built-in read-only memory (ROM) on which secure, immutable elements of the firmware (e.g., the bootloader) can be installed. Together, the TPM and the secure memory (including the bootloader) comprise the meter's TCB. A graphical representation of the meter's main components can be found in \Cref{fig:system_model} in \Cref{sec:system_model}.

\textit{Blockchains. }
A \emph{blockchain} is a data structure that allows a network of mutually distrusting parties to maintain a tamper-resistant digital ledger. Elementary write operations called \emph{transactions} are created by nodes within the network and broadcast to the other nodes. 
Transactions are grouped into \emph{blocks}. Starting from the \textit{genesis} block, the blockchain is extended through a \emph{consensus protocol} in which nodes are consecutively selected to propose new blocks. When a node is selected, it chooses a group of not-yet-included transactions, decides on their ordering, and executes them sequentially. 
 The transactions are then added to a block alongside metadata such as a timestamp and the hash of a previous block. Finally, the block is transmitted across the network. 
 
A chain can be constructed from any block to the genesis block through the previous-block hashes. This protects against data tampering: if as little as a single bit is changed in a block, then its hash is different and the chain is broken. Blockchains can be \emph{permissionless}, i.e., any node is allowed to propose blocks if they meet certain requirements such as computational power, or \emph{permissioned}, i.e., only a fixed, known set of nodes can propose blocks \cite{dinh2017blockbench}. In this work, we focus on permissioned (or \textit{private}) blockchains, and the transactions are always calls to functions of \emph{smart contracts}, which are pieces of software uploaded to the blockchain.

\textit{Cryptographic Primitives. }
Several security mechanisms discussed in this paper, including the FLBI framework of \Cref{sec:architecture}, use digital signatures. A digital signature scheme \cite{goldwasser1988digital} consists of the \textit{key generation}, \textit{signing}, and \textit{verification} algorithms $\keygen$, $\sign$, and $\verify$ as follows:
\begin{itemize}
\item $(\sk, \pk) \leftarrow \keygen(1^{\lambda})$ takes a security parameter $\lambda$ as input, and outputs a secret key $\sk$ and a corresponding public key $\pk$. 
\item $\sigma \leftarrow\sign(\sk, \msg)$ takes a secret key $\sk$ and a message $\msg$ as input, and outputs a valid signature $\sigma$ on $\msg$.
\item $v \leftarrow \verify(\pk,\msg,\sigma)$ takes a public key $\pk$, message $\msg$, and signature $\sigma$ as input, and outputs a value $v \in \{0,1\}$ such that if $v=1$, then $\sigma$ is a valid signature of $\msg$ under $\pk$ with overwhelming probability, and $v=0$ if and only if $\sigma$ is an invalid signature of $\msg$ under $\pk$. 
\end{itemize}
We focus on two main types of digital signature schemes. The first type relies on a single entity with a single secret key.
 The second type uses \textit{threshold signatures} \cite{gennaro2018fast}. In the latter approach, $n$ parties agree on a threshold $t$ and participate in the $\keygen$ protocol. The output of $\keygen$ consists of a single shared public key $\pk$ and a set $S = \{\sk_1,\ldots,\sk_n\}$ with a secret key $\sk_i$ for each participant $i$. The secret keys used for $\sign$ can be any set $S^* \subseteq S$ with at least $t$ members -- i.e., at least $t$ out of the original $n$ members must collaborate to generate the signature. However, the signature that is produced is not necessarily more complex than one produced by the first type of signature scheme. Similarly, $\verify$ is not necessarily more complex for the second type of scheme than for the first. For the scheme used in our implementation \cite{gennaro2018fast}, the resulting signature is an ECDSA signature \cite{johnson2001elliptic} that is as easy to verify in a smart meter's bootloader as a single-party signature.

\section{System Model and Requirements}
\label{sec:system_model}

\subsection{System Model: Entities} Our system model consists of the following entities:

\textit{Input signal. } This represents the physical process that the meter is measuring, e.g., power flow on a line, or water or gas in a pipeline.

\textit{Smart meter. } The meter itself consists of \textit{trusted} and \textit{untrusted} components. Trusted components include the TPM, the code in the ROM (including the bootloader), and the secure memory.
The untrusted parts include the sensor, the device firmware, and the components controlled by the firmware, i.e., the RAM and flash memory, the communications module, and the processing unit. Other smart meter components -- e.g., LCD displays or the power supply -- do not affect data integrity and are therefore not included in the model.

\textit{MDMS server. } This is a server run by the utility company. It operates a Meter Data Management System (MDMS) \cite{KumarLBPDM19,skopik2015smart,stellios2018survey,sridhar2011cyber}, and is hence responsible for the storing and processing of measurement data.

\textit{Data storage. } This is where the measurement data is stored. The storage can be seen as part of the MDMS system.

\textit{OEM server. } This is run by the \textit{Original Equipment Manufacturer} (OEM), i.e., the company that produces and maintains the hardware and firmware of the meters. The OEM is responsible for firmware updates, which are sent from the OEM server.

\textit{DCUs. } The meters are connected to \textit{Data Concentrator Units} (DCUs) over a Local Area Network (LAN). The relationship between the meters and the DCUs is many-to-one, and the relationship between the DCUs and the company server is also many-to-one. DCUs enhance the scalability of the system by reducing the number of devices that the company server needs to poll to retrieve measurements.
 DCUs are typically located at a physically secure location in a building complex or distribution substation  \cite{skopik2015smart,toledo2013smart}.

\textit{WAN. } The DCUs communicate with the utility company servers over a Wide Area Network (WAN), e.g., the Internet. Firmware updates are sent over the same network.

\subsection{System Initialization}

The data flows in the system -- measurements and firmware binaries -- can be digitally signed. In particular, the measurements are signed by the smart meter's TPM, and the firmware binaries are signed by the OEM server. The digital signature scheme used by these two entities need not be the same: the TPM's algorithms are denoted by $\tpmkeygen$, $\tpmsign$, and $\tpmverf$, and the OEM server's algorithms by $\oemkeygen$, $\oemsign$, and $\oemverf$. We assume that $\tpmkeygen$ is executed prior to the production of the meters, that the resulting secret key $\tpmsk$ is hardcoded in the TPM, and that the public key $\tpmpk$ is shared with the utility company. We also assume that $\oemkeygen$ has been executed prior to the meters' production, that the resulting secret key $\oemsk$ is stored on the OEM's server, and that the public key $\oempk$ is either stored in the meter firmware or in the ROM (we discuss this in more detail in Section~\ref{sec:existing_countermeasures}).

In our model, the meter's flash memory has \textit{two partitions}: the first partition stores the current firmware version, and the second a new binary that is checked after the next reboot.

\subsection{Measurement Flow}

After initialization, the normal flow of measurement data consists of steps 1-8 as depicted in \Cref{fig:system_model}.

\textit{1. } In the first step, the sensor generates a measurement $x$ based on the physical input signal and sends it to the processing unit.

\textit{2 \& 3. } Next, the processing unit sends $x$ to the TPM, which produces a signature $\sigma_x = \tpmsign(\tpmsk,m)$. The TPM sends $\sigma_x$ back to the processing unit.

\textit{4 -- 7. } The processing unit then forwards $x$ and $\sigma_x$ to the meter's communications module, which sends this data to the company server via the DCU and the WAN.

\textit{8. } Finally, the company server receives both $x$ and $\sigma_x$ and determines $v_x = \tpmverf(\tpmpk, x, \sigma_x)$. If $v_x=1$, then $\sigma_x$ is valid and $x$ is sent to the storage. For additional security, $\sigma_x$ can be stored as well, as we discuss in Section~\ref{sec:existing_countermeasures}.

\subsection{Firmware Flow}

The normal flow of firmware updates consists of steps i-v as depicted in \Cref{fig:system_model}.

\textit{i -- iii. } In the first step, the OEM creates a new firmware binary $b$ and produces a signature $\sigma_b = \oemsign(\oemsk, b)$. This is then sent to the smart meter via the network and the DCU  -- in practice, the update would typically be sent to the utility companies' servers first, but this is considered part of sending it through the network, i.e., step i. Note that sending firmware updates via the DCUs relieves the OEM or company server from sending a sizable firmware binary to potentially millions of meters through direct connections.

\textit{iv. } After it has been received by the meter, $b$ and $\sigma_b$ are copied to the second partition of the meter's flash memory, and a reboot is triggered.

\textit{v. } After the reboot, the bootloader checks whether a firmware binary and signature are available on the second partition and whether it is the same as the first partition's binary. If not, it computes $v_b = \oemverf(\oempk, b, \sigma_b)$, and if $v_b = 1$, then $b$ and $\sigma_b$ are written to the first partition. If  $v_b = 0$, or if the firmware version is the same on both partitions, then the bootloader loads the first partition's binary and signature -- if the signature is valid, it boots the firmware from this partition, and terminates the boot process otherwise.

\subsection{System Requirements}
\label{sec:requirements}

An AMI should meet the following security goals, which it shares with IoT sensor networks in general: 
\begin{itemize}
\item {\bf Data integrity}: each stored measurement corresponds to the state of the physical input signal at the time of recording.
\item {\bf Firmware integrity}: the binary that is loaded by the meter's bootloader corresponds to the latest version that has been produced by the OEM and received by the meter.
\end{itemize}
To have practical relevance, a design that meets our security requirement should also satisfy the following system goals:
\begin{itemize}
\item {\bf Scalability:} it can handle the load generated by a real-world AMI consisting of a large number of meters.  
\item {\bf Deployability:} it incurs minimal changes to existing AMIs. 
\end{itemize}

\section{Threat Model}
\label{sec:threat_model}

In the previous section, we presented our model of the normal flow of measurement and firmware data. In our model, four types of data are sent between the entities: the measurement $x$, the firmware binary $b$, and the corresponding signatures $\sigma_x$ and $\sigma_b$. The \textit{goal} of the attacker is to violate data (or firmware) integrity by altering $x$ (or $b$) before or at the end of the flow. The attack surface consists of the $7$ different entities in our model.
For brevity, we denote the attack against entity $i$ by $A_i$ for $i =1,\ldots,7$.

\textit{Input signal \& sensor ($A_1$). } If the attacker has direct access to the meter, then it can be possible for her to tamper with the input signal without touching the meter itself. For example, a power sensor may be distorted by holding a magnet close to it, or high-wattage devices can be disconnected from the
meter’s circuit and connected directly to the grid \cite{mclaughlin2013multi}. In this attack, the integrity of the measurement $x$ is altered at the source, and $\sigma_x$ is altered because $x$ is changed before it is signed. Tampering with the meter's sensor has the same effect.

\textit{Meter firmware ($A_2$). } An attacker with direct access to the meter may be able to replace the device firmware directly. For example, the meter may have an input channel (such as a USB port) that allows the attacker to flash the firmware. There may also be a zero-day vulnerability in the current firmware that can be exploited, or it may be possible to physically replace the flash memory that contains the binary. This attack allows the attacker to alter the version of $b$ on either memory partition, but not $\sigma_b$ because this would require the OEM's secret key. If the attacker controls the device firmware or the communication between the components in the device, then it is possible to alter $x$, and additionally $\sigma_x$ if $x$ is changed before it is signed.

\textit{MDMS server ($A_3$). } The computer systems of the utility company can be compromised, for example through hacking, spear-phishing
 \cite{ukraine2016analysis}, or malicious insiders \cite{homoliak2019insight}.  This may allow the attacker to obtain passwords that give
 access to authorized communication channels, or that give write access to the data stored on the MDMS servers. This enables the attacker to alter $x$, but not $\sigma_x$ because the TPMs of the devices are not affected.
 Access to the secure communication channel may also enable the attacker to push malware to the meters via the DCUs if updates pass by the MDMS servers on their way from the OEM servers to the meters. As such, the attacker may send an altered $b$ to the meters via the DCUs, but not a matching $\sigma_b$ because that would require $\oemsk$, the OEM's secret key. If there is a log of the firmware update history, then this can also be overwritten by the attacker.
 
\textit{Data storage ($A_4$). } If the attacker gains access to the data storage, then she can alter the historical measurements $x_1,x_2,\ldots,x_n$. However, she will not be able to alter the historical signatures $\sigma_{x\,1},\sigma_{x\,2},\ldots,\sigma_{x\,n}$ since this would require access to the TPM keys.

\textit{OEM Server ($A_5$). } The computer systems of the OEM are vulnerable to the same types of attack as the MDMS servers. Compromising
 the OEM servers may allow an attacker to obtain the key $\oemsk$ that allows her to sign a malicious firmware binary $b$, or to
 send such a $b$ to the utility company via the WAN. Hence, this enables her to alter both $b$ and $\sigma_b$. An attacker may even be able to retrieve TPM secret keys if these have been stored, or during the production of the meters. This would allow the attacker to alter measurement signatures if $A_5$ is combined with other attacks.
 
\textit{DCUs ($A_6$). } Compromising the firmware on the DCU may allow the attacker to alter both $x$ and $b$. However, it does not allow for the signatures $\sigma_x$ and $\sigma_b$ to be altered.

\textit{WAN ($A_7$). } Although the communication between the DCUs and OEM and MDMS servers can be made secure using SSL/TLS, this can be compromised by an attacker, e.g., through the use of a forged X.509 certificate. The effects are the same as for the DCUs ($A_6$): the attacker can alter $x$ and $b$, but not $\sigma_x$ and $\sigma_b$.

\section{Limitations of Existing Countermeasures}
\label{sec:existing_countermeasures}

\subsection{Existing Defense Mechanisms}
\label{sec:existing_overview}

In this section, we discuss four existing defense strategies for data integrity attacks that are suitable in the context of AMI data and firmware integrity.  
We identify which combinations of defense mechanisms satisfy the requirements of Section~\ref{sec:requirements}, and discuss their limitations. 

\textit{Signature Checks ($D_1$, $D_2$, $D_3$, and $D_4$).}
To prevent arbitrary firmware from being installed, the meters check whether any new firmware binary $b$ has a correct signature $\sigma_b$
\cite{certicom2018,eetimes2019}. In practice, the OEM's signature can be checked immediately before an update
\cite{eetimes2019}, or after every reboot by the bootloader which we assume to be stored on the smart meter's ROM \cite{schaller2014lightweight,certicom2018}. In our model, we have assumed the latter. The public key that is used to verify the signature can be stored on the (untrusted) flash memory, or on the (trusted) ROM. We label the first case as $D_1$, and the second case -- i.e., a hard-coded public key -- as $D_2$.

Similarly, the utility company verifies measurement signatures. The company can verify the signatures of incoming measurements ($D_3$) before storing them, and if the signatures are stored then the signatures of historical measurements can be verified as well ($D_4$).

\textit{Remote Attestation ($D_5$).}
In this mechanism, either the TPM \cite{kil2009remote,sailer2004design,sridhar2011cyber} or a cumulative attestation kernel \cite{lemay2012cumulative} provides remote attestation, or is able to sign messages with a version-dependent private key. 
This allows the MDMS server to detect firmware tampering by querying the meter's current firmware version and comparing it to a history of firmware updates stored on the server.
We include specification-based intrusion detection 
\cite{berthier2011specification,jokar2011specification}
as a variation of this defense mechanism.

\textit{Anomaly Detection ($D_6$).}
Anomaly detection is a means of detecting data
tampering. We consider two categories: in the first, the recent measurement patterns of a single meter are
compared to the measurements of the other meters \cite{golomb2018ciota}. In the second, the recent
measurements of a meter are compared to the same meter's historical usage pattern
\cite{laptev2015generic,feng2019systematic,jokar2015electricity,hangal2002tracking}. 
Anomaly detection can also be used to detect firmware tampering, but this requires data -- e.g., memory jump sequences \cite{golomb2018ciota} -- outside our system model. The accuracy of anomaly detection methods depends on factors such as the size and recency of the training dataset, and the generality of the class of anomalies that is considered, but a full discussion of these aspects is beyond the scope of this paper.

\textit{Blockchains ($D_7$). }
Blockchains have the property that as long as (at least) a majority of the nodes (weighted by consensus power) has not been compromised, then attempts to change blockchain data are instantly detectable. This property can be applied to secure the integrity of historical measurements, and of the firmware update log on the MDMS server. Several blockchain-based approaches have been proposed in the literature for this purpose \cite{maw2019ics,LiangWLZD19,lee2017blockchain,YohanL18,HuYLY19}. A blockchain can store aggregated, periodically uploaded metadata (e.g., a hash of a large batch of measurements, or a hash of the entire dataset), or the (encrypted) data itself. In either case, this data is replicated on each node that participates in the blockchain.

\subsection{Limitations}
\label{sec:limitations}

\Cref{tab:threat_model_summary} summarizes the relationship between the different data types and the attacks and defense mechanisms. For measurements and the corresponding signatures, we distinguish between incoming measurements (``new'') and historical (``log'') measurements that have already been stored. For the binaries, we distinguish between the log of historical firmware versions on the MDMS servers (``log''), the current (``curr.'') firmware version on the meter's first partition, and the new firmware binary on the meter's second partition (``new''). Some defense mechanisms require the integrity of certain data types -- e.g., remote attestation ($D_5$) requires an accurate log of firmware updates ($b$ (log)), whereas anomaly detection ($D_6$) requires accurate training data ($x$ (log)).

\renewcommand{\arraystretch}{1}
\begin{table}[hbt]
\centering
\caption{Summary of the relationship between the various data types and the attacks and defense mechanisms.}
\label{tab:threat_model_summary}
\begin{tabular}{c|ccc}
data type & threatened by & protected by & required by \\ \toprule
 $x$ (log) 			& $A_4$ 						& $D_4$, $D_7$ 		& $D_6$ \\
 $x$ (new) 			& $A_{1-4}$, $A_{6}$, $A_{7}$	& $D_3$, $D_6$		& \\ \midrule
 $\sigma_{x}$ (log) & $A_5$ 						& 					& $D_4$ \\
 $\sigma_{x}$ (new) & $A_{1}$, $A_{2}$, $A_5$ 		& 					& $D_3$ \\ \midrule
 $b$ (log) 			& $A_3$ 						& $D_7$ 			& $D_5$ \\
 $b$ (curr.)		& $A_2$ 						& $D_5$ 			& $D_1$ \\
 $b$ (new) 			& $A_{2}$, $A_{3}$, $A_{5-7}$  	& $D_{1}$, $D_2$			& \\ \midrule
 $\sigma_b$ (log) 	& $A_4$ 						& 					& \\
 $\sigma_b$ (curr.) & $A_5$ 						& 					& \\
 $\sigma_b$ (new) 	& $A_5$ 						& 					& $D_{1}$, $D_{2}$ \\ \bottomrule
\end{tabular}
\end{table}
\renewcommand{\arraystretch}{1.5}

We can now identify the combinations of defense mechanisms that satisfy the requirements of data integrity and firmware integrity as defined in Section~\ref{sec:requirements}. In particular, \textit{data integrity} is satisfied if integrity violations in new $x$ can be detected in all combinations of attack vectors in our threat model -- similarly, \textit{firmware integrity} is guaranteed if integrity violations in new $b$ can always be detected in our threat model. 
In the following, we discuss which combinations of mechanisms satisfy data integrity and firmware integrity -- however, we find that all of these have limitations.

\textit{Firmware Integrity. } signature checks can detect firmware integrity violations. However,  performing these checks in the current firmware version ($D_1$) is insecure unless the current firmware version would be part of the TCB. However, this would \textit{significantly increase the TCB size}. This limitation can partially be addressed by periodically checking the current firmware version through remote attestation ($D_5$) -- however, this is slow and depends on the integrity of the log of firmware versions on the MDMS servers, which itself would need a blockchain ($D_7$) to be secure. Furthermore, $D_1$ is unable to detect integrity violations if the OEM's key has been stolen and the attacker has signed the compromised binary ($A_5$).

Performing the checks inside the bootloader ($D_2$) addresses the first limitation by keeping the TCB small without needing remote attestation for the current firmware version. However, a na\"ive implementation of $D_2$ is still unable to detect attacks when the OEM's key is stolen -- in fact, in this case, the meters are unable to fully recover without a factory recall as the OEM's key is stored in the secure memory, and such a recall would violate our requirement of deployability. In Section~\ref{sec:architecture}, we show how to address this limitation of $D_2$. By using multiparty signatures and evolving \textit{signature chains}, the meter is able to recover from stolen OEM keys without a recall. Furthermore, a cooldown period for the multiparty signatures gives the OEM a time window to detect that its key has been stolen. Verifying the multiparty signatures inside the meter's TCB is non-trivial -- however, we show how this can be done efficiently using threshold signatures.

\textit{Data Integrity. } incoming measurements $x$ can be protected by a blockchain-based log ($D_7$) combined with anomaly detection ($D_6$). This protects $x$ from all attacks, including tampering with the sensor or input signal ($A_1$). If only hashes are logged on the blockchain, then this mechanism can detect any form of tampering with $x$. However, this combination of mechanisms would not be able to recover from data tampering: it would conclude that the training data has been tampered with, but the attack would still render all forms of anomaly detection using this dataset invalid in the future. This can be addressed by storing all (encrypted) data on the blockchain. However, this requires that large quantities of data are stored on the blockchain, so  \textit{scalability} is a limitation. 
In the next section, we address this limitation through a two-layer architecture.

\section{The FLBI Framework}
\label{sec:architecture}

\begin{figure}[tb]
    \centering
    \includegraphics[width=0.45\textwidth]{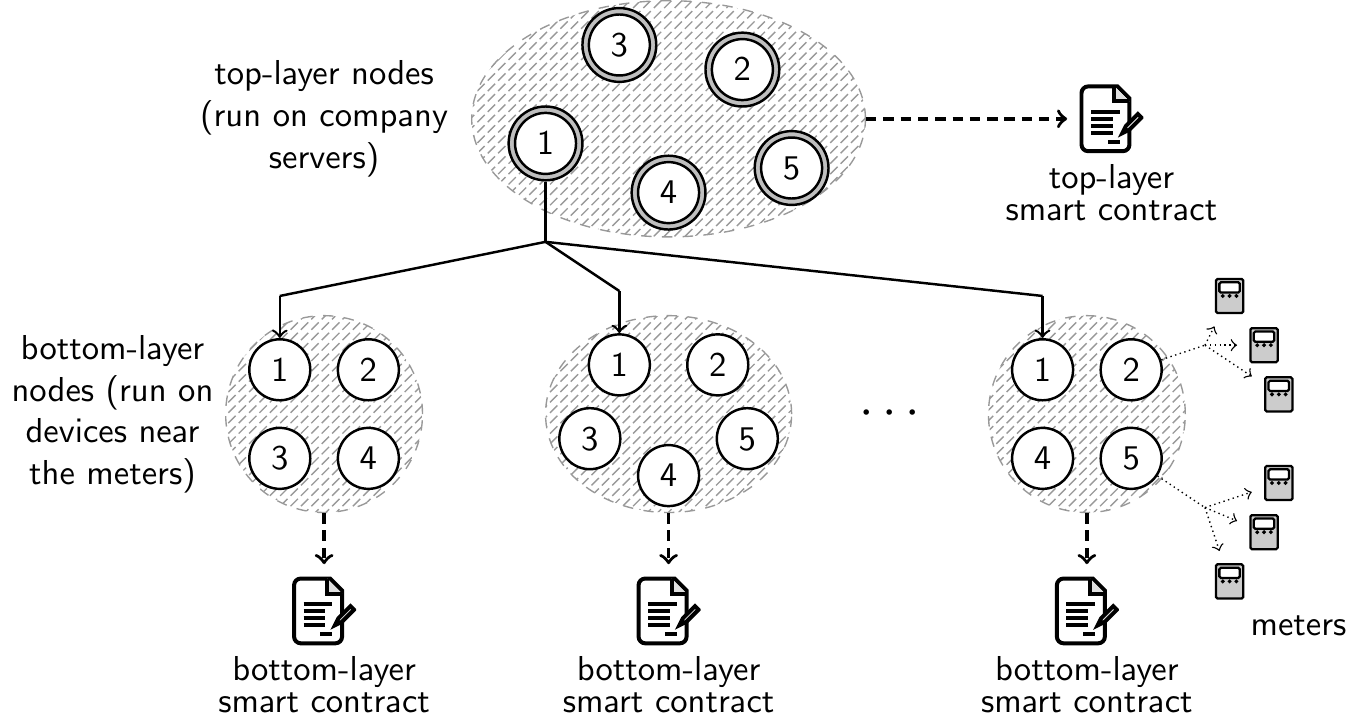}
    \caption{Design of the FLBI.}
    \label{fig:architecture}
\end{figure}

In this section, we present the FLBI framework, which allows a consortium of operators and other interested parties to address the limitations identified in \Cref{sec:limitations}. 
In the following, we first discuss our notation and the entities that participate in the FLBI. We then briefly discuss the programming code and the end-to-end protocols of the core processes: measurement logging, firmware updates, and changes to the consortium. A graphical representation of the two-layer blockchain framework can be found in \Cref{fig:architecture}. In this example, a consortium consisting of 5 members operates a top-layer blockchain. There are also many bottom-layer blockchains, which are operated by nodes that run on low-power devices owned by the consortium members. Each bottom-layer node can have a large number of meters assigned to it.

\subsection{Notation \& Assumptions}

In the following, we assume that each top-layer node can sign messages. As such, its digital signature scheme consists of the algorithms $\hlnkeygen$, $\hlnsign$, $\hlnverf$. During initialization each top-layer node $c$ runs $\hlnkeygen$, stores the resulting secret key $\hlnsk_c$, and shares $\hlnpk_c$ with the other top-layer nodes. Each meter $j$'s TPM has secret key $\tpmsk_j$ and public key $\tpmpk_j$.

We use the following notation for the threshold signature scheme: $\threskeygen$, $\thressign$ and $\thresverf$. 
We denote the set of members (represented as integers) after the $i$th membership update by $C_i$. Let $n_i = |C_i|$ be the size of $C_i$, and $f_i$ the largest integer such that $n_i \geq 3 f_i + 1 $. The $\threskeygen$ algorithm is run during initialization and results in a shared public key $\pk_0$ and set of private keys $S_0 = \{\sk_{0,1},\ldots,\sk_{0,n}\}$. After the $i$th membership update, $i>0$, let $\pk_i$ be the shared public key, and $\sk_{i,j}$ be the secret key of member $j$. Let $S_i$ be the set of secret keys after the $i$th update, i.e., $S_i = \{\sk_{i,j}\}_{j \in C_i}$. We assume that all the public keys and private keys are unique (i.e., even if $C_i = C_j$ for some $i<j$, we expect the members to generate new keys and not reuse the old ones). 

\subsection{Entities}

\textit{Consortium Members. }
The \textit{consortium} is the group of companies and organizations that are involved in operating the blockchains. The consortium must include utility companies who operate meters and who add data to the bottom-layer blockchains, and OEMs who add information about firmware updates to the top-layer blockchain.
\footnote{If there are OEMs who are responsible for some added meters but who do not join the consortium, then information about their firmware updates can be added by the utility companies as well.} 
Other interested
parties such as regulators and consumer organizations can operate nodes on both levels, even if they do not
add measurements or firmware updates.

For the FLBI framework to be sustainable in the long term, it must be able to support periodic changes to the set of consortium members. As such, the consortium's membership set is \textit{dynamic}. For any update, if $n = 3f+1$ members are in the current set, then at least $2f+1$ need to agree to the update through a digitally signed vote. Therefore, it is advisable that $n$ is large enough to recover from multiple failures (e.g., permanent losses of signing keys). Meanwhile, the consortium should avoid including frivolous members who cannot be trusted to respond to proposed updates, as the protocol can tolerate at most $f$ abstentions when the membership set is updated. As such, a balance must be struck between including too few or too many members. In the following, we assume that $n$ is roughly equal to $16$.

\textit{Top-Layer Blockchain. }
The top-layer blockchain provides access control -- i.e., the public keys of the consortium members, of the top-layer nodes, and the meters. It also maintains a log of firmware updates and stores the latest shared public key $\pk_i$. 
A single smart contract called the \textit{top-layer contract} implements all functionalities of the top-layer blockchain. Each consortium member runs a single top-layer node on a highly available online computer, e.g., an MDMS server. The computer that runs the top-layer node also runs an application called the \textit{top-layer client} that provides an interface to create smart contract transactions, e.g., consortium updates or new firmware updates. The top-layer client also transmits information to the bottom-layer nodes that belong to the same company, e.g., to push firmware binaries and public key updates. This mapping between top-layer nodes to bottom-layer nodes is one-to-many. To produce threshold signatures, the top-layer clients have the ability to communicate with each other to interactively execute the $\thressign$ protocol.

To defend against blockchain-level attacks, the top-layer clients also execute a protocol to ensure that no bottom-layer blockchain $l$ becomes dominated by a single consortium member $c$. In particular, when any $c$ creates a proposal to add a new node to $l$, then the other clients compute for \textit{all} bottom-layer blockchains how many nodes are controlled by $c$ in that blockchain. The high-layer nodes vote to reject the proposal if $l$ is not one of the bottom-layer blockchains on which
$c$ is represented the \textit{least}. 
That is: let $n_{cl}$ be the number of nodes controlled by $c$ in bottom-layer blockchain $l$, $L$ the set of bottom-layer blockchains, and $N_{l}$ the total number of nodes in blockchain $l \in L$. Let $\phi_{cl} = n_{cl}/N_{l}$ be the fraction of nodes controlled by $c$ in $l$. 
In our protocol, the high-level clients vote to reject adding a new node operated by consortium member $c^*$ to
$l^*$ unless $\phi_{c^*l^*} = \min_{l \in L} \phi_{c^*l}$. Hence, the protocol seeks to
enforce that $\phi_{cl} < \frac{1}{3}$ for each $c$ and $l$.  

\textit{Bottom-Layer Blockchains. }
The bottom-layer blockchains provide secure measurement logging. 
Each bottom-layer blockchain operates a single smart contract called the \textit{bottom-layer contract}.
Each smart contract stores a different set of measurement data, so there is no need for interaction between different bottom-layer blockchains. The bottom-layer nodes are run on devices that are situated
close to the meters, e.g., on the DCUs or on devices placed in the same secure location. The computational power of the nodes is assumed to be at least
comparable to that of a Raspberry Pi 4. 

Each bottom-layer blockchain node runs an application called the \textit{bottom-layer client} that listens for messages from the high-level client. If it receives a public key update, it will propose the same update to its blockchain if it has not been proposed by another node already. If it is not the first to propose, it will vote to support updates that match the ones that it has received and ignore other updates. 

\subsubsection*{Meters}

Finally, to address the challenge of recoverability, we update the bootloader of the smart meter to support updates to the public key stored in its ROM through signature chains.  

\subsection{Smart Contracts}

Due to space limitations, we do not discuss the detailed implementations of the smart contract functions and the clients in the main text.  However, a summary of the smart contract functions is given in \Cref{tab:smart_contract_functions}, and the code used in the experiments can be found at \github.

\renewcommand{\arraystretch}{1}
\begin{table}[htp]
\caption{Overview of the smart contract functions}
\vspace{-0.1cm}
\label{tab:smart_contract_functions}
\begin{small}
\begin{center}
\begin{tabular}{cc|c}
smart contract & functionality & function name \\ \toprule
\multirow{4}{*}{top layer} & \multirow{3}{*}{firmware updates} & $\putfw$ \\
 & & $\vetofw$ \\
 & & $\checkfw$ \\ \cline{2-3}
 & access control & $\updateadminkey$ \\ \midrule
\multirow{2}{*}{both layers} & \multirow{2}{*}{access control} & $\proposeput$ \\
 & & $\voteput$ \\ \midrule
bottom layer & measurement logs & $\putlog$ \\ \bottomrule
\end{tabular}
\end{center}
\end{small}
\vspace{-0.6cm}
\end{table}%
\renewcommand{\arraystretch}{1.5}

\subsection{End-to-End Protocols}

We present a high-level overview of the end-to-end protocols for measurement logging, firmware updates, and consortium membership updates below.

For measurement logging, each measurement $x$ is sent from the meter to its bottom-layer node, which encrypts it and uploads it to the bottom-layer blockchain through a call to the $\putlog$ function from Table~\ref{tab:smart_contract_functions}. 

For firmware updates, a hash of the binary $b$ is uploaded by the OEM server to the blockchain, and after the end of the cooldown period, the OEM asks a group of at least $2f+1$ other consortium members to produce a threshold signature for the binary hash. The binary and signature are sent to the meter, which hashes the binary, checks the signature for the hash, and if successful boots the new firmware.

\begin{figure}[htb]
    \centering
    \includegraphics[width=0.43\textwidth]{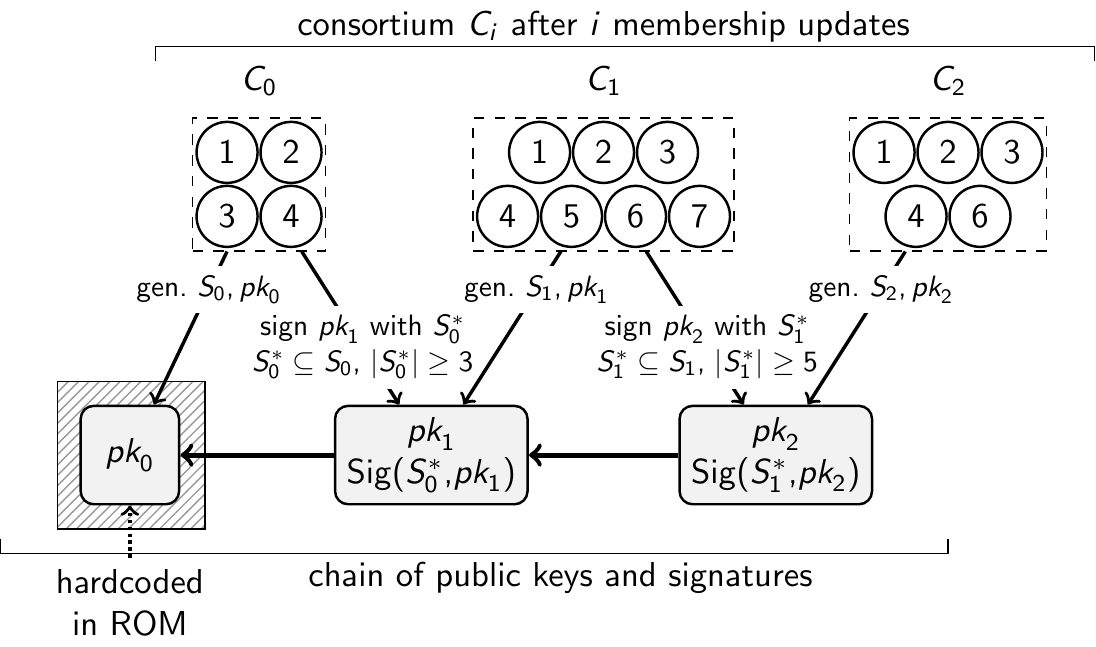}
    \caption{Evolution of the key/signature chain as members join and leave the consortium.}
    \label{fig:signature_chain}
\end{figure}

To reflect consortium membership updates, each firmware update requires that the OEM produces a chain $(\pk_0,\sigma_1,\pk_1,\ldots,\sigma_n,\pk_n)$ of alternating public keys and signatures. Here, $\pk_0$ is the initial public key that is hard-coded in the meter's ROM, $\pk_i$ is the public key after the $i$th membership update, and $\sigma_i = \thressign(S^*_{i-1}, \pk_i)$ is the threshold signature for $\pk_{i+1}$ produced by a group $S^*_{i-1}$ that consists of at least $2f_{i+1}+1$ members of the ${i-1}$th consortium. This process is depicted in Figure~\ref{fig:signature_chain}. The chain is sent to the meter alongside the firmware binary. Next, the bootloader verifies the chain through the repeated verification of ECDSA signatures -- if successful, the meter updates the consortium's public key to the last public key in the chain. This ensures that the bootloader uses the current consortium's public key before checking the signature of the new binary.

\subsection{Analysis: Security Against Blockchain Attacks}
\label{sec:analysis}

As discussed in \Cref{sec:limitations}, blockchains have security requirements that do not fit into the system model. We provide an informal analysis of two notable attacks in this section.

\textit{Minority Attacks in the Bottom-Layer Blockchains. }
In the FLBI framework, the bottom layer consists of $L$ blockchains that can each be attacked independently. Each bottom-layer blockchain
uses PBFT \cite{castro1999practical} for consensus, which is secure only if at most
$f$ out of the $n \geq 3f+1$ of the nodes are compromised. It is a challenge to ensure that this threshold holds in every
bottom-layer blockchain.  
Most of the layer-two and sharding proposals in the literature \cite{dang2019towards,kokoris2018omniledger} use
periodic reconfiguration of the nodes among the bottom-layer blockchains to achieve a diverse or unpredictable mixture of nodes.
Our setting allows for a simpler solution with a stricter guarantee. The reason is that the nodes in different bottom-layer blockchains are not
independent, because one entity (e.g., a utility company) typically controls many nodes across different bottom-layer blockchains. In \Cref{sec:architecture}, we have described a protocol in which clients vote to reject proposals to add a bottom-layer node to a chain unless it is the chain on which the proposing entity is represented the \textit{least}. As a result, no single entity can concentrate its nodes in a single bottom-layer blockchain, hence violating the $n \geq 3f+1$ requirement if it is compromised by an attacker.

\textit{Network and timing attacks. }
Our network assumption is partially synchronous -- i.e., partitions are eventually resolved, and nodes' messages can eventually reach the rest of the network. However, communication from the meters to the outside world may be disrupted, due to
attacks such as radio jamming, DoS, or other network-level attacks. Such attacks quickly
become evident when a meter's readings stop appearing on the log. 
However, temporary network disruption may impact the security of the top-layer blockchain operations. More
specifically, both firmware updates and changes to the access control list use a cooldown
period during which vetoes can be cast. This period
is
important for preventing compromised keys to sign malicious firmware. In particular, during this period the
signing key can be revoked, rendering the proposed firmware invalid. If this cooldown period is implemented
using a local clock, then the adversary can disrupt the network until the cooldown period is over, preventing any veto
from making it onto the blockchain. To mitigate this, we implement the cooldown duration in terms of the number of
blocks. For instance, a cooldown period of $T$ blocks means that a reject vote must be cast within $T$ blocks
from the block in which the proposal is recorded. In other words, after $T$ blocks, the proposal is considered to be
valid. Our solution works with the assumption that a transaction sent at the same time as block $b$ will get
included in the blockchain at block $b+T$ at the latest.  

A practical choice for $T$ is to multiply $\tau$, the desired cool-down period length in minutes, by the average number of blocks added to the blockchain per minute. However, an adversary who controls one or more companies' top-layer nodes can influence the speed at which new blocks are added, e.g., by spamming empty blocks. This threat is mitigated by using a round-robin consensus protocol to select block proposers. An adversary who controls $f$ out of $n$ nodes and who proposes new blocks in negligible time can at worst reduce the duration of the cool-down period to $\frac{n-f}{n} \tau$ minutes, as $n-f$ out of every $n$ blocks will be proposed by non-compromised nodes. A round-robin protocol also protects against \textit{censorship} attacks, in which the adversary refuses to include blacklisted transactions -- e.g., a transaction that revokes a key. Since one non-compromised node is selected at least every $n$ blocks, it will include the transaction and this block will be committed if at least $2f+1$ non-compromised nodes vote to support it. We do require a form of transaction prioritization, as an adversary could otherwise attempt to flood the \textit{mempool} -- i.e., the list of pending transactions -- of the non-compromised nodes with spurious transactions. As such, we give key revocations and vetoes a higher priority at the level of the consensus protocol than other votes and limit the ability of nodes to create this type of transaction.

\section{Performance Evaluation}
\label{sec:evaluation}

To evaluate the practical performance of the FLBI framework, we have fully implemented it on top of both private Ethereum and Hyperledger Fabric (``Fabric'' for brevity). In the following, we first describe the experimental setup in detail and then discuss the experimental results on transaction throughput and latency. 
We note that we do not have an appropriate baseline for an experimental comparison, because the other blockchain approaches for data integrity that we are aware of do not present realistic end-to-end experiments.

\begin{figure*}[ht]
\centering
\subfloat[][]{\includegraphics[width=0.2\textwidth]{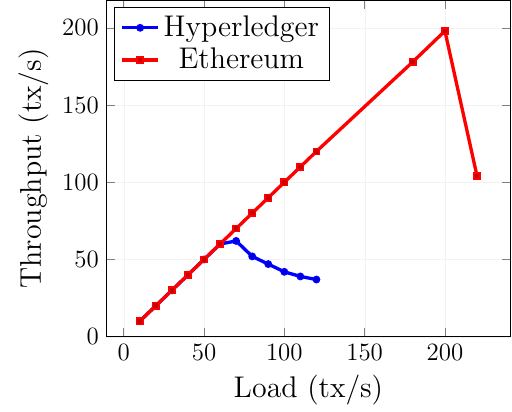}\label{fig:load_throughput}}
\subfloat[][]{\includegraphics[width=0.2\textwidth]{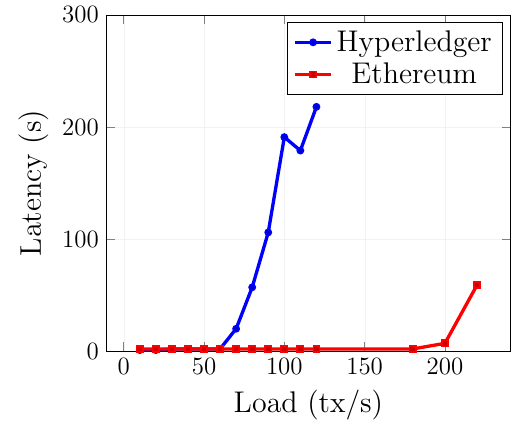}\label{fig:load_latency}}
\subfloat[][]{\includegraphics[width=0.2\textwidth]{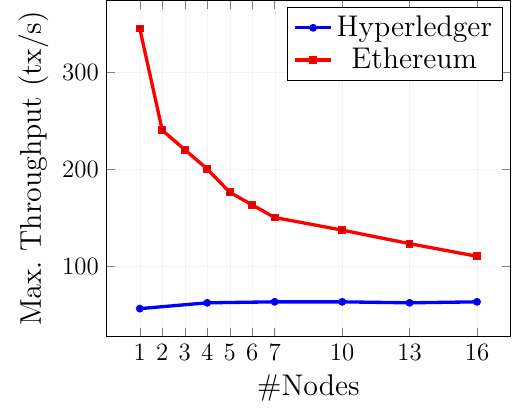}\label{fig:peak_throughput}}
\subfloat[][]{\includegraphics[width=0.2\textwidth]{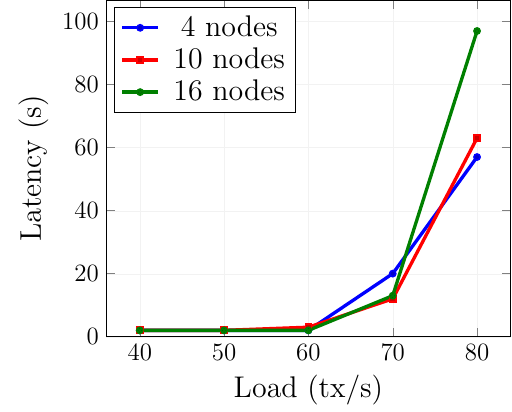}\label{fig:threshold_latency}}
\subfloat[][]{\includegraphics[width=0.2\textwidth]{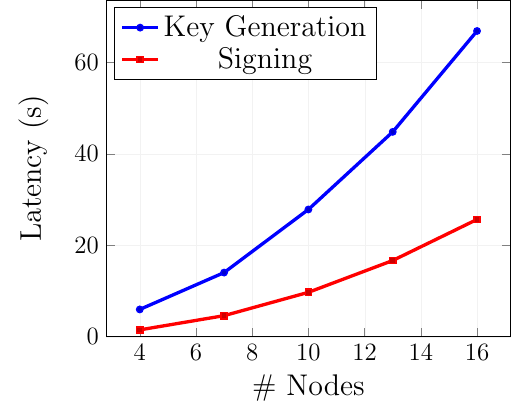}\label{fig:latency_networksize}}
\caption{(a) Throughput vs.\ load on 4 nodes for logging. (b) Latency vs.\ load on 4 nodes. (c) Max.\ throughput vs.\ number of nodes. (d) Latency vs.\ load for $4$-$16$ nodes in Hyperledger. (e) Latency of threshold key generation and signing algorithms.}
\vspace{-0.15cm}
\end{figure*}

\subsection{Experimental Setup} 
 We use Fabric
version 0.6, which supports the PBFT consensus protocol and has higher performance than version
1.4~\cite{dinh2017blockbench}. For Ethereum, we use \texttt{geth} version 1.9. The smart contracts are written in
Go and Solidity. We collect data generated by 6 Ampohub B1 smart meters
that run in our local testbed, and replay it through a client running on a laptop. The data fields
include: meter identifier, supplier identifier, metadata, power, voltage, and frequency readings.

Our setup consists of two blockchains. The top-layer blockchain consists of $4$ nodes, such that each node has 16 GB
RAM and an 8-core CPU. The bottom-layer blockchain consists of up to $16$ nodes, each node runs on a Raspberry Pi
(RPI) model 4B, each with 4 GB RAM and a 64 GB SD card. The Raspberry Pis, which have limited computational capabilities, represent the bottom-layer blockchain nodes. The Pis are connected through a NETGEAR 24-Port
gigabit Ethernet unmanaged switch. Each Pi runs the Raspbian Buster operating system. To evaluate the firmware update protocol, we use a Linkit smart 7688 development board. In total, we have added \textit{259 lines of code} to the TCB -- in particular, to {U-Boot} \cite{uboot}, which acts as the bootloader for the Linkit. 

\subsection{Blockchain performance} 
Figures~\ref{fig:load_throughput} and
\ref{fig:load_latency} compare the performance of Hyperledger and Ethereum using $4$ nodes for
measurement logging (i.e., using the $\putlog$ function). Each measurement is cryptographically hashed, timestamped, and signed before
it is sent to the blockchain. The peak throughputs of Hyperledger and Ethereum are $60$ and
$200$ transactions per second (txs), respectively. The reason that Ethereum performs better is that it
parallelizes the signature verifications for multiple transactions within the same block, as opposed to
verifying them sequentially.  If each meter produces a reading every 30 minutes (as in \cite{straitstimes2019power}), then a throughput of $60$ txs means that $108,000$ meters can
be supported in each bottom-layer blockchain. In this case, ten bottom-layer blockchains would be able
to support more than a million meters.  Figures~\ref{fig:peak_throughput}~and~\ref{fig:threshold_latency} show peak throughput and latency
with a varying number of nodes. We can see that the throughput does not change for Hyperledger, suggesting
that consensus is not the bottleneck. However, the throughput degrades significantly for Ethereum, meaning that
the consensus protocol becomes a bottleneck as each bottom-layer network grows in size.  

\begin{figure}[!tbp]
  \centering
  \subfloat[Message size and Execution cost]{\includegraphics[scale=0.45]{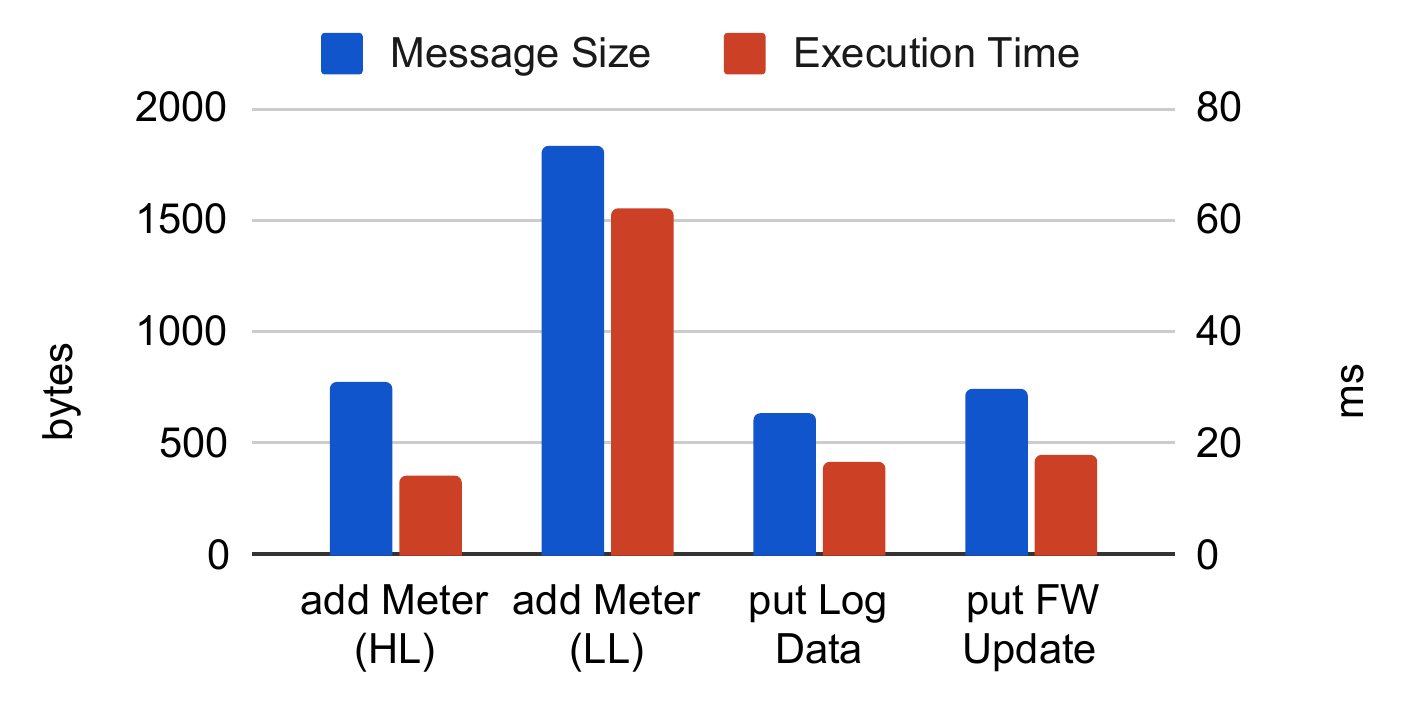}\label{fig:exec_cost}}
  \hfill
  \subfloat[CPU (\%) for {\tt putLogData} calls]{\includegraphics[scale=0.5]{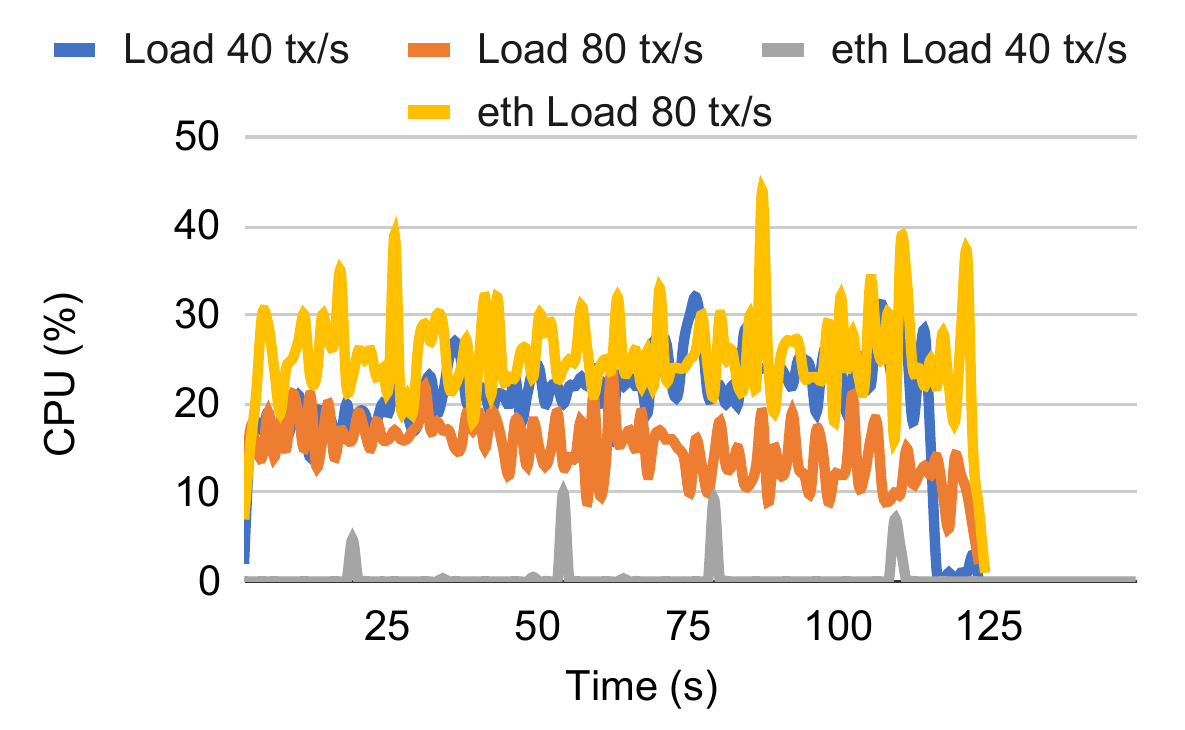}\label{fig:CPUutlization40_80}}
\caption{Resource utilization with 16 nodes.}
\vspace{-0.1in}
\label{fig:util}
\end{figure}

To better understand this performance, we measure the cost at the blockchain node. For Hyperledger, the
consensus latency is $20ms$ on average, and remains unchanged for varying network sizes.
Figure~\ref{fig:exec_cost} shows the execution costs for different types of transactions in Hyperledger. It
can be seen that the cost for normal transactions is in the order of tens of milliseconds, which explains the
low overall throughput. Figure~\ref{fig:util} shows the CPU and I/O utilization at the node. It can be seen
that none of the resources are fully utilized. These results suggest that the overall throughput can be further
improved by adding concurrency to transaction execution, to ensure that more transactions can be executed simultaneously.

We have also tested the latency of the two main threshold signature algorithms -- $\threskeygen$ and $\thressign$ -- for different consortium sizes. For our experiments, we have used a library (\mbox{{https://github.com/binance-chain/tss-lib}}) that implements the protocol by Gennaro \& Goldfeder \cite{gennaro2018fast}.  We have performed our experiments on the device that also runs the top-layer blockchain nodes. The results are depicted in \Cref{fig:latency_networksize}. Key generation is the most expensive step, as it takes more than a minute for 16 consortium members. However, this step has to be performed only when the consortium is changed. Signing takes slightly over 25 seconds for 16 members. This step is performed only when a new firmware binary is signed. By contrast, the cost of verifying the signature is the same as verifying any ECDSA signature - on the Linkit device, this took less than a tenth of a second. From this, we can conclude that we have successfully shifted the computational burden away from the low-power device's TCB. 

\begin{figure}
    \centering
    \includegraphics[width=\linewidth]{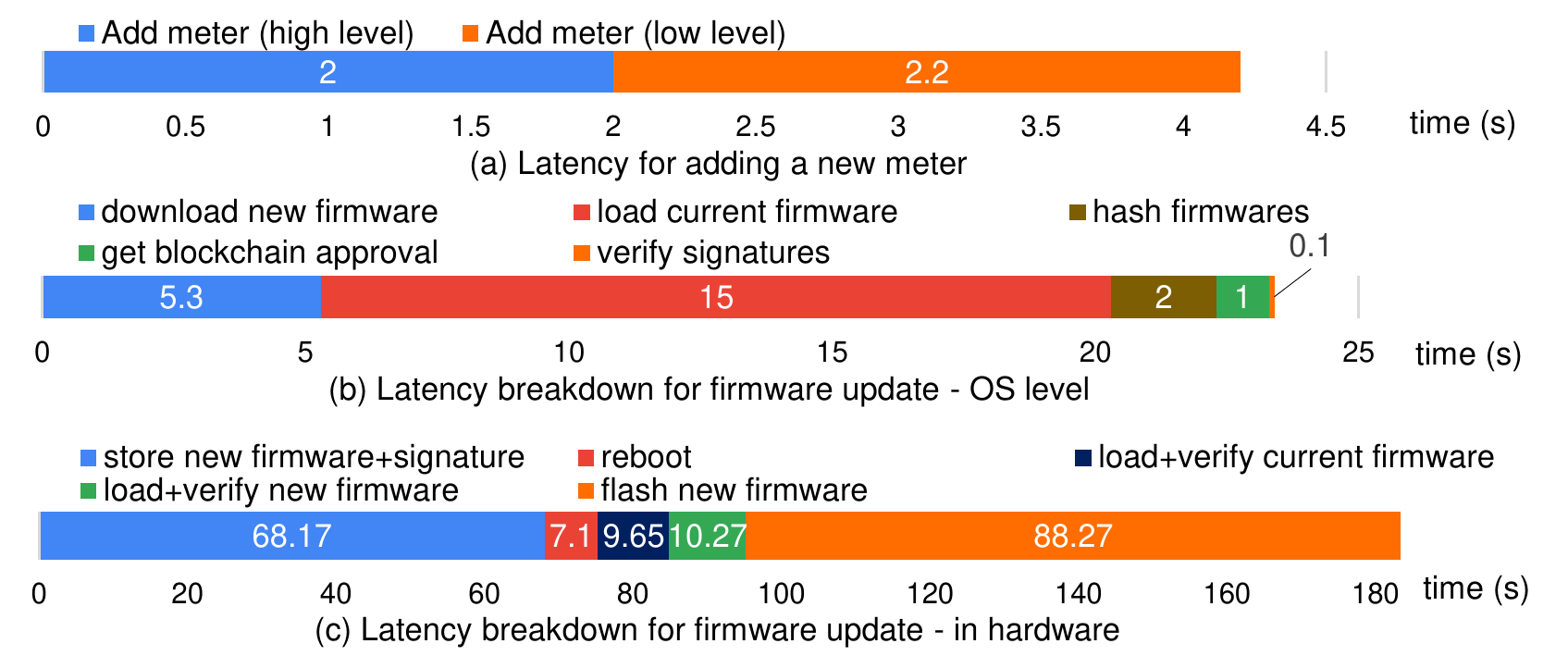}
    \caption{Latency breakdown for adding new meters and firmware updates}
    \vspace{-0.15cm}
    \label{fig:fw_update_duration_1}
\end{figure}

\subsection{End-to-End latency}
Figure~\ref{fig:fw_update_duration_1}a shows the end-to-end latency for adding a new meter in the FLBI framework. The cost
consists of adding a public key to the top-layer blockchain, approving the new key, requesting a
proof, and adding the public key to the bottom-layer blockchain. The entire process takes roughly $4s$, and the
main cost comes from adding the key to each blockchain.  

Figure~\ref{fig:fw_update_duration_1}b and \ref{fig:fw_update_duration_1}c show the latency for updating the new firmware on the device. There are two main procedures: 1) downloading and verifying the new firmware on the blockchain network at the OS level, and 2) verifying and flashing it at the hardware level. The cost
includes: registering the firmware to the blockchain, downloading the firmware to the device, getting the new
hash from the top-layer blockchain, verifying the signature, and finally flashing the new firmware. 
The major cost comes from writing the new firmware to the flash memory partitions, which amounts to about $156s$ in total. The protocol
on the OS level takes less than $25s$.

\section{Related Work}
\label{sec:related_work}

\textit{Generic Mechanisms for Data Integrity. }
Generic mechanisms that protect data integrity at any layer of an IoT sensor network (device, network, or service) can be found in the literature on IoTs, mobile networks, and distributed systems.
To protect
the device firmware, existing solutions include code signing \cite{kim2018broken}, deterministic builds
\cite{de2014challenges}, remote attestation \cite{kil2009remote}, anomaly detection
\cite{hangal2002tracking,feng2019systematic,golomb2018ciota,jokar2015electricity}, and blockchain solutions
such as CHAINIAC \cite{nikitin2017chainiac}. To protect measurements stored at the utility company servers, existing solutions include
certificate transparency~\cite{Laurie14}, witness cosigning \cite{syta2016keeping}, and
blockchains~\cite{maw2019ics,Al-BassamM18,DorriKJG19,dorri2017towards,dorri2017blockchain}. 

If seen as individuals, each of these approaches falls short in the context of security-critical sensor networks, and especially AMIs, for three
reasons. 
First, AMIs involve multiple stakeholders for whom privacy is important. In particular, measurement
logs can reveal valuable information about the user, and the firmware typically contains business secrets. The
latter rules out approaches that would worsen privacy by relying on multiple parties evaluating the source code or attesting to the
correspondence between the firmware's source code and binary~\cite{nikitin2017chainiac,Al-BassamM18}. Second,
individual mechanisms have a narrow scope that fails to consider the interdependence between attacks against the
AMI's different layers \cite{wang2019xlf} or components. For example, anomaly detection techniques that aim to identify malicious firmware
by identifying suspicious measurements \cite{jokar2015electricity} will fail if the measurement logs have been
overwritten to include abnormal behavior. Third, due to its critical nature, AMIs require strong guarantees
against Byzantine adversaries \cite{castro1999practical}, which is not met by solutions such as certificate
transparency whose security depends on auditing frequency and network conditions.

In \Cref{sec:existing_countermeasures}, we have discussed how the second obstacle can be overcome for remote attestation, anomaly detection, and blockchain-based solutions within a unifying framework.


\textit{Blockchain Mechanisms for Data Integrity. }
Several blockchain mechanisms have been proposed in the literature that address data integrity at the device layer, but which are not suitable in our context. Liang et al.\ \cite{LiangWLZD19} propose a proof-of-work blockchain that is operated by the meters and stores encrypted measurement data. However, proof-of-work is vulnerable to 51\% attacks by powerful miners \cite{homoliak2020security} and is known to waste computational resources, which is a major obstacle on low-resource devices such as meters. They also do not present an end-to-end implementation of their system, but present simulation results instead.

Lee and Lee \cite{lee2017blockchain} propose an approach for secure IoT firmware updates in which a number of high-power devices maintain a proof-of-work blockchain that stores a history of firmware versions. Low-power devices periodically query a high-level node for the latest firmware version. If a new firmware version is detected, it is sent to the requesting node in a peer-to-peer manner. Although this method shares some similarities with the one presented in \Cref{sec:architecture}, this work does not address the question of how to recover the firmware once it has been corrupted, as the software client that runs on the meter and interacts with the blockchain is itself part of the firmware, and therefore outside the TCB. Furthermore, it does not consider data integrity beyond the firmware, and does not present an implementation or experiments. Yohan and Lo \cite{YohanL18} present a similar approach that shares the same limitations. Hu et al.\ \cite{HuYLY19} propose several optimizations compared to Lee and Lee \cite{lee2017blockchain} and present an evaluation of a private Ethereum chain run on an emulated Raspberry Pi (through Qemu-Pi). However, they also do not consider recoverability or attacks against data integrity beyond the device layer.

\textit{Two-Layer Blockchains. }
In \Cref{sec:architecture}, we present a blockchain framework that uses a two-layer protocol to achieve scalability. The challenge of speeding up blockchains through multi-layer protocols is an active research area. Proposals generally fall into either of two categories: \textit{layer-two} mechanisms, which rely on a top-layer chain that connects a large number of largely independent blockchains \cite{gudgeon2020sok}, and \textit{sharding} mechanisms, which partition the entire blockchain workload at the level of the consensus protocol \cite{kokoris2018omniledger,elhindi2019blockchaindb}. As the bottom-layer blockchains in our framework do not communicate, our approach is part of the former category. Since there are, to the best of knowledge, no off-the-shelf layer-two protocols available that are suitable for large-scale data logging on the bottom layers, we have implemented and tested our framework in Hyperledger Fabric and private Ethereum.

\section{Conclusions \& Discussion}
\label{sec:conclusions}

In this paper, we have presented a lightweight blockchain framework for securing the integrity of data and device firmware in IoT sensor networks. We have presented a comprehensive threat model that includes the interaction between attacks on the network's different entities. Our threat model allows us to identify which defense mechanisms satisfy the security goals of \textit{measurement data integrity} and \textit{firmware integrity} while maintaining real-world scalability and deployability. 
We have identified three main limitations of existing designs and addressed these through 1) a two-layer blockchain design, 2) analysis of blockchain-specific attacks, and 3) a novel application of threshold signatures that are verified in the TCB of the meters.
 We have implemented our solution using smart contracts in Ethereum and Hyperledger Fabric, and have performed stress tests that show its applicability in practice. We have made the code of the experiments publicly available. In contrast to firmware code, the public availability of the smart contract code implies that it can be audited -- therefore, it is less likely to contain zero-day vulnerabilities.

The FLBI framework is applicable in a broad range of IoT settings -- the main requirements are that 1) each device collects data independently from other devices, 2) the devices have TPM support, and 3) a consortium of interested parties can be found to maintain a healthy balance of participants in each bottom-layer blockchain (i.e., no entity controls more than $f$ out of $3f+1$ nodes). Regarding the latter requirement, we note that a blockchain solution in which a single entity controls all the peers offers no security beyond that of a centralized infrastructure. However, even in applications in which the measurements collected by different operators (e.g., healthcare providers) are fully independent, security can be enhanced if they use bottom-layer blockchains to share encrypted data. 

With a throughput of 60-200 transactions per second and 32 bytes of meaningful payload (the hash) per transaction, each bottom-layer blockchain has an estimated throughput of approximately 2-6 kB/s. This may be sufficient for (encrypted) AMI or patient monitoring data, but less so for raw security camera footage. However, in the latter case, it may be sufficient to store the output of a machine learning algorithm applied to the raw video footage on the bottom-layer blockchains. This requires that the bottom-layer devices have the capacity to compute such output, but this is unlikely to be more computationally expensive than maintaining a blockchain node. 

\section*{Acknowledgments}
This research / project is supported by the National Research Foundation, Singapore, under its National Satellite of Excellence Programme ``Design Science and Technology for Secure Critical Infrastructure'' (Award Number: NSoE\_DeST-SCI2019-0009). Any opinions, findings and conclusions or recommendations expressed in this material are those of the author(s) and do not reflect the views of National Research Foundation, Singapore.

\bibliographystyle{IEEEtran}
\bibliography{ref}

\end{document}